%% file: main.tex
\documentclass{webofc}
\usepackage[varg]{txfonts}   
\usepackage[utf8]{inputenc}
\usepackage[T1]{fontenc}
\usepackage{standalone}

\usepackage{amsmath}
\usepackage{url}
\usepackage{multirow}
\usepackage{colortbl}
\usepackage[colorlinks=true,linkcolor=firebrick,citecolor=darkgreen,urlcolor=darkblue]{hyperref}
\usepackage{mathptmx} 
\usepackage{enumerate}
\usepackage{lineno}
\usepackage{xspace}
\usepackage{multicol}
\definecolor{darkred}{rgb}{0.5,0,0}
\definecolor{darkblue}{rgb}{0,0,0.5}
\definecolor{firebrick}{rgb}{0.75,0.125,0.125}
\definecolor{darkgreen}{rgb}{0,0.5,0}

\newcommand{\NASixtyOne}{\mbox{NA61/SHINE}\xspace}

\newcommand{\pp}{\mbox{\textit{p}+\textit{p}}\xspace}

\newcommand{\MeV}{\mbox{Me\kern-0.1em V}\xspace}
\newcommand{\GeV}{\mbox{Ge\kern-0.1em V}\xspace}
\newcommand{\A}{\textit{A}\xspace}
\newcommand{\GeVc}{\mbox{Ge\kern-0.1em V\kern-0.15em /\kern-0.05em\textit{c}}\xspace}
\newcommand{\MeVc}{\mbox{Me\kern-0.1em V\kern-0.15em /\kern-0.05em\textit{c}}\xspace}

\newcommand{\AGeVc}{\A~\GeVc}
\newcommand{\Ar}{\textsuperscript{40}Ar\xspace}
\newcommand{\Sc}{\textsuperscript{45}Sc\xspace}
\newcommand{\ArSc}{\mbox{\Ar~+~\Sc}\xspace}
\newcommand{\Pb}{\textsuperscript{208}Pb\xspace}
\newcommand{\PbT}{Pb\xspace}
\newcommand{\PbPb}{\mbox{\Pb~+~\PbT}\xspace}
\newcommand{\dedx}{\mbox{\ensuremath{\textrm{d}E\!/\!\textrm{d}x}}\xspace}

\begin{document}
\title{Search for the critical point of strongly interacting matter}
\subtitle{( Intermittency analysis by \NASixtyOne at CERN SPS )}

%

\author{\firstname{Haradhan} \lastname{Adhikary for the \NASixtyOne collaboration}\fnsep\thanks{\email{haradhan.adhikary@cern.ch}}
}

\institute{Jan Kochanowski University, Kielce, Poland}

\abstract{%
The  existence  and  location  of  the  QCD  critical  point  is  an  object  of both experimental and theoretical studies.  The comprehensive data collected by \NASixtyOne during a two-dimensional scan in beam momentum (13\A-150\AGeVc) and system size (\pp, \textit{p}+Pb , Be+Be, Ar+Sc, Xe+La, Pb+Pb) allows for a systematic search  for  the  critical  point  –  a  search  for  a  non-monotonic  dependence  of  various correlation and fluctuation observables on collision energy and size of colliding nuclei. In particular, fluctuations  of  particle  number  in  transverse  momentum  space  are studied.  They are quantified by measuring the scaled factorial moments of multiplicity distribution.

}
\maketitle
\input{sections/introduction.tex}

\input{sections/na61shine.tex}

\input{sections/particle-identification}
\input{sections/methodology.tex}

\input{sections/results.tex}

\input{sections/models.tex}

\input{sections/pion-intermittency}

\input{sections/summary.tex}
\section{Acknowledgements}
This work is supported by the National Science Centre, Poland under grant no. 2018/30/A/ST2/0026. 
\bibliography{include/NA61Reference}

\end{document}

%% file: sections/introduction.tex
\section{Introduction}

In this proceeding, experimental results on scaled factorial moments of proton and negatively charged hadron multiplicity distribution in central \PbPb collisions at 13\AGeVc, 30\AGeVc, and  \ArSc collisions at 13\A-150\AGeVc are presented.
The measurements were performed by the multi-purpose
\NASixtyOne~\cite{Abgrall:2014xwa} experiment at the CERN Super Proton Synchrotron (SPS).
As part of the strong interactions program, the \NASixtyOne studies properties of the onset of deconfinement and searches for the critical point of the strongly interacting matter. Within this program, a two-dimensional scan of collision energy and the size of colliding nuclei was performed~\cite{Aduszkiewicz:2642286}. The reported results concern the search for the critical point (CP).

%% file: sections/na61shine.tex
\section{The \NASixtyOne detector}
\label{sec:detector}


The \NASixtyOne detector is a large-acceptance hadron spectrometer
situated in the North Area H2 beam-line of the CERN SPS~\cite{Abgrall:2014xwa}. The main components of the detection system are four large-volume Time Projection Chambers (TPCs).
This setup allows for precise momentum reconstruction and identification of charged particles. The high-resolution forward hadron calorimeter, the Projectile Spectator Detector (PSD), is used to determine the collision's centrality by measuring the energy of projectile spectators, i.e. non-interacting projectile nucleons.

%% file: sections/particle-identification.tex
\section{Particle identification}
\label{sec:identification}
\begin{figure}[!ht]
    \centering
    \includegraphics[width=.5\textwidth]{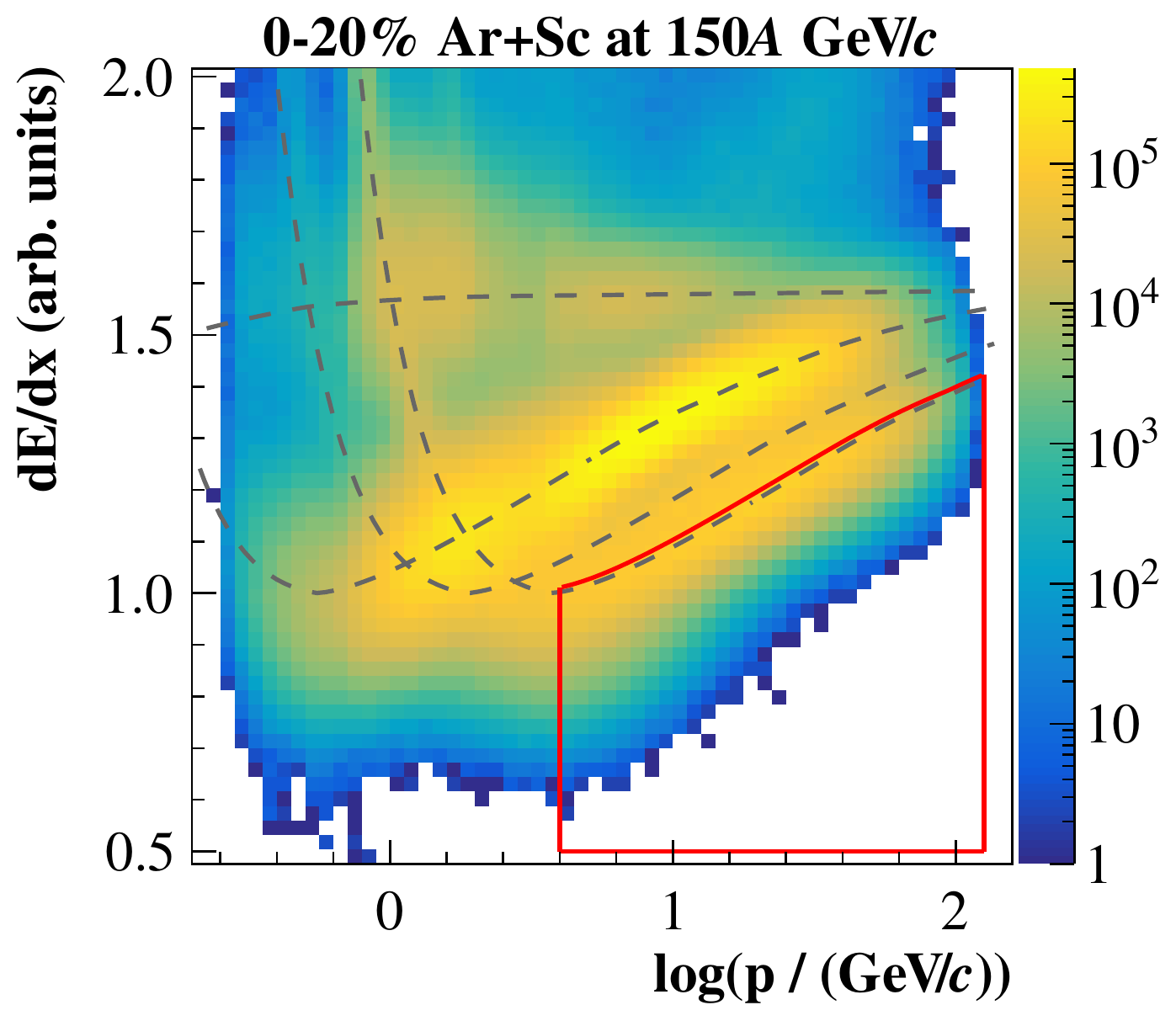}
    \caption{
        Energy loss vs total momentum of positively charged particles measured with the \NASixtyOne
        Time Projection Chambers (TPCs) in central \ArSc collisions at 150\AGeVc.
        Dashed lines picture the nominal Bethe-Bloch values for protons, kaons, pions and electrons.
        The cut to select proton candidates is marked with a red line.  
    }
    \label{fig:dEdx}
\end{figure}

Based on the ionization energy loss (\dedx) in TPCs (Fig.~\ref{fig:dEdx}) charged particles with different momenta are identified as protons or negatively charged hadrons.  As an example of proton selection, the \dedx distribution for positive particles is shown in Fig.~\ref{fig:dEdx} and the selected region is marked with the red line. On average about 60\% of all protons are selected for the analysis. Kaons contaminate this sample by less than 4\%.
\\
A similar procedure was used to select negatively charged hadrons (h$^{-}$) by removing electrons using dE/dx cut.

%% file: sections/methodology.tex
\section{The critical point search methodology}
\label{sec:sfm}
The goal of the analysis is to search for the critical point of the strongly interacting matter
by measuring the scaled factorial moments for mid-rapidity particles using cumulative variables of transverse momenta (section:~\ref{sec:sfm_cumulative}) and statistically independent points (section:~\ref{sec:independentpoints}).

\subsection{Critical point and intermittency in heavy-ion collisions}
\label{sec:sfm_CP}

A second-order phase transition leads to the divergence of the correlation length ($\xi$).
The infinite system becomes scale-invariant with the particle correlation function having the power-law form, which appears as intermittent behavior of particle multiplicity fluctuations~\cite{Wosiek:1988}.
The intermittent multiplicity fluctuations~\cite{Bialas:1985jb}  were
discussed as the signal of CP by Satz~\cite{Satz:1989vj}, Antoniou et al.~\cite{Antoniou:1990vf} and Bialas, Hwa~\cite{Bialas:1990xd}.
This initiated experimental studies of the structure of the phase transition region via analyses of
particle multiplicity fluctuations using scaled factorial moments (SFMs) ~\cite{NA49:2012ebu}. Later, additional measures of
fluctuations were also proposed as probes of the critical
behavior ~\cite{Stephanov:1998dy,Stephanov:1999zu}. \\
The \NASixtyOne experiment
has performed a systematic scan of collision energy and system size. The new measurements may answer
the question about the nature of the transition region and, in particular,
whether the critical point of strongly interacting matter exists and where it is located.
The scaled factorial moments $F_r(\delta)$~\cite{Bialas:1985jb} of order $r$ are defined as
\begin{equation}
  F_r(\delta) = \frac
    {\bigg<{\displaystyle{\frac{1}{M}\sum_{i=1}^{M}} N_i(N_i-1)...(N_i-r+1) }\bigg>}
    {\bigg<{\displaystyle{\frac{1}{M}\sum_{i=1}^{M}} N_i }\bigg>^r }\:,
  \label{eq:scaled-factorial-moments}
\end{equation}
where $\delta$ ($=\frac{\Delta}{M^{1/D}}$) is the size of each of the $M$ subdivision intervals of the
selected interval $\Delta$ (two-dimensional bins in the transverse-momentum plane), $N_{i}$ is the particle multiplicity in a given interval, and angle brackets denote averaging over the analysed events.
If the system at freeze-out is close to the CP, then the system is a simple fractal
and $F_r(\delta)$ follows a power-law dependence:
\begin{equation}
\label{eq:cp_1}
  F_{r}(\delta) = F_{r}(\Delta) \cdot (\Delta / \delta)^{D\cdot\phi_{r}}~.
\end{equation}
Moreover, the exponent (intermittency index) $\phi_{r}$ obeys the relation:
\begin{equation}
\label{eq:cp_2}
  \phi_{r} = ( r - 1 )/D \cdot d_{r}\:,
\end{equation}
where $D$ denotes the dimensionality of the system and the anomalous fractal
dimension $d_{r}$ is independent of $r$~\cite{Bialas:1990xd}.
Such behaviour is the analogue of critical opalescence in conventional
matter~\cite{Antoniou:2006zb}.\\
The QCD-inspired
considerations~\cite{Antoniou:2000ms,Stephanov:2004wx} suggest that the critical behaviour can be observed by measuring the fluctuations of the number of protons based on
the assumption that the critical fluctuations are transferred to the net-baryon density; the net-baryon density may therefore serve as an order parameter of the phase transition ~\cite{Fukushima:2010bq, Hatta:2002sj, Stephanov:2004wx,
Antoniou:2008vv,Karsch:2010ck,Skokov:2010uh,Morita:2012kt}. Thus, such fluctuations are expected to be present in the net-proton number and the proton and anti-proton numbers separately~\cite{Hatta:2003wn}. For protons,
$\phi_2 = 5/6$~\cite{Antoniou:2006zb} is expected.\\

\subsection{Cumulative transformation}
\label{sec:sfm_cumulative}
\begin{figure}[!ht]
    \centering
    \includegraphics[height=5cm]{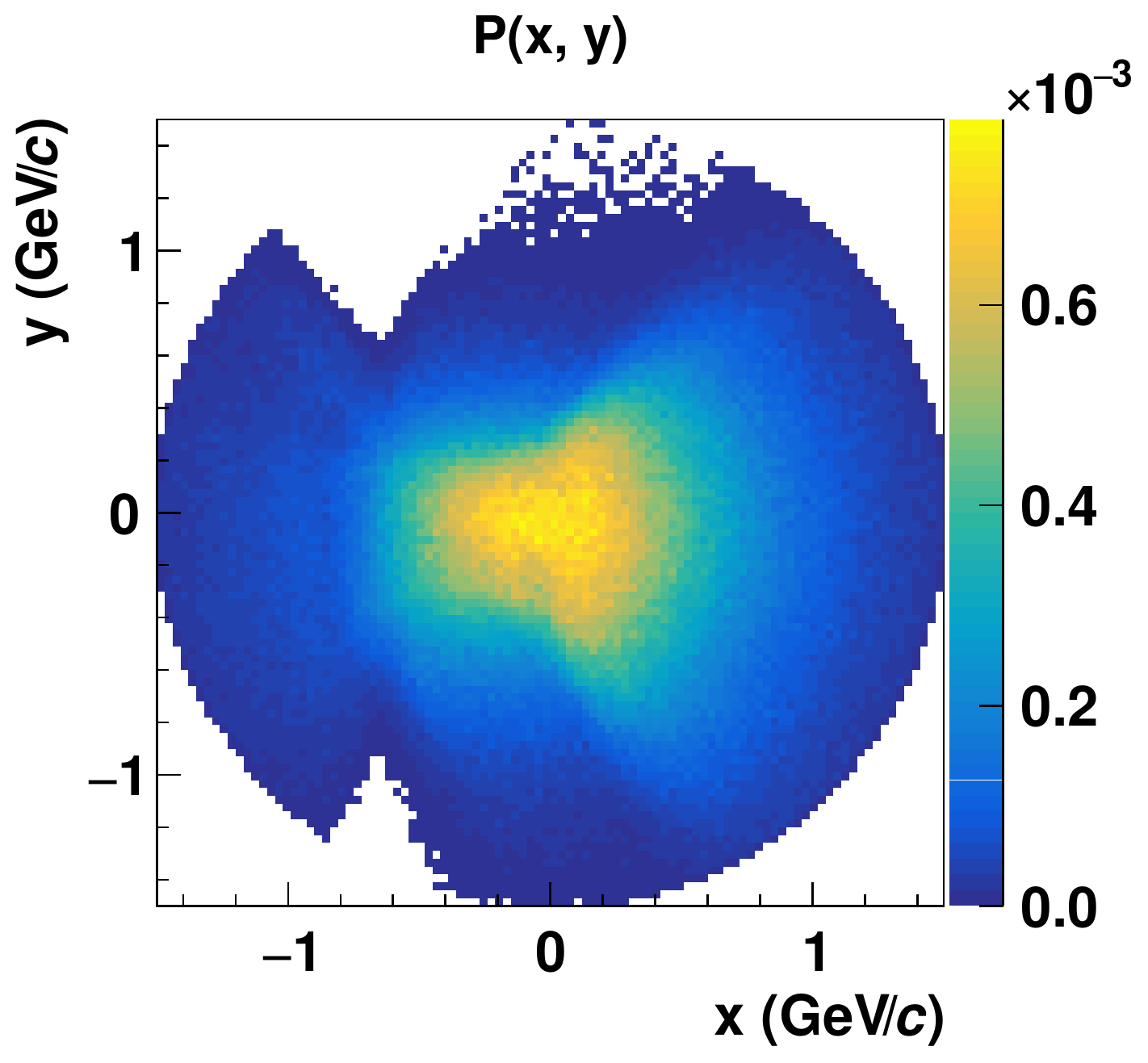}\quad
    \includegraphics[height=5cm]{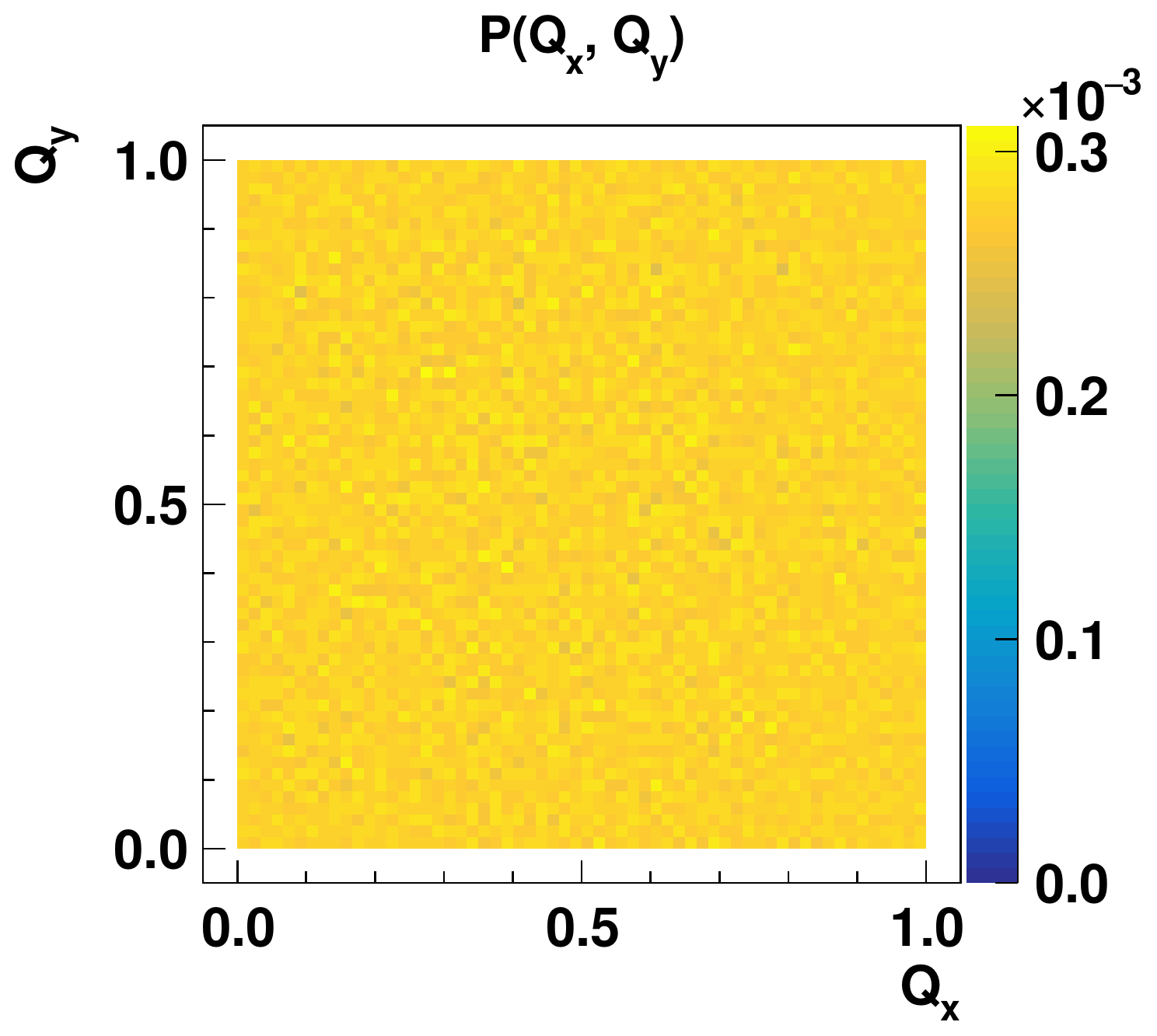}
    \caption{
        Example of the effect of the cumulative transformation of transverse momentum components,
        $p_{x}$ and $p_{y}$ (denoted as $x$ and $y$), selected for proton
        intermittency analysis of the \NASixtyOne \ArSc at 150\AGeVc data.
        Distribution before (\emph{left}) and after (\emph{right}) the cumulative transformation is presented.
    }
    \label{fig:cumulative}
\end{figure}

SFMs are sensitive to the shape of the single-particle distribution. The momentum distribution is generally non-uniform and this dependence may bias the signal of critical fluctuations. To remove this dependence one can use cumulative variables~\cite{Bialas:1990dk}. At the same time, it was verified~\cite{Samanta:2021dxq} that the transformation preserves the critical
behaviour given by Eq.~\ref{eq:cp_1}.
By construction, particle density in the cumulative variables is uniformly distributed.  For one-dimensional single-particle distribution in \textit{x}, the cumulative variable is
$$
    Q_{x} = \int\limits_{a}^{x} \rho(x)dx \Bigg/ \int\limits_{a}^{b} \rho(x)dx~,
$$
where $a$ and $b$ are lower and upper limits of the variable $x$.
For a two-dimensional distribution $\rho(x,y)$ and a given $x$ the transformation reads
$$
    Q_{y}(x) = \int\limits_{a}^{y} \rho(x,y)dy \Bigg/ \int\limits_{a}^{b} \rho(x,y)dy.
$$
The distribution of $Q_{x}$ and $Q_{y}$ is a uniform
from 0 to 1 (Fig.~\ref{fig:cumulative}).
\\

\subsection{Statistically-independent data points}
\label{sec:independentpoints}
The intermittency analysis gives the dependence of scaled factorial moments on
the number of subdivisions of transverse-momentum or cumulative transverse-momentum space.
Here statistically independent data subsets were used to obtain results for each subdivision number.
In this case, the results for different subdivision numbers are statistically uncorrelated. Thus the full relevant information needed to interpret the results (the central value and variance) is easy to present graphically and use in the statistical tests.
However, the procedure decreases the number of events used to calculate each data point.

%% file: sections/results.tex
\section{Intermittency analysis results}
\label{sec:results}
This section presents the results on scaled factorial moments  
(Eq.~\ref{eq:scaled-factorial-moments}) of selected particles (\textit{p} or $h^{-}$) produced by strong and electromagnetic processes in central (0-10\%) \PbPb at 13\AGeVc, central (0-10\%) \PbPb at 30\AGeVc and central (0-10\%) \ArSc collisions at 13\A-75\AGeVc and central (0-20\%) \ArSc at 150\AGeVc collisions. 
The results are shown as a function of  the number of subdivisions of
cumulative transverse-momentum space -- the so-called intermittency analysis. The independent data sets are used for subdivision.
The typical acceptance for intermittency analysis is given in ~\cite{na61AccMapProtinIntermittencyArSc150}.

\subsection{Results for protons in central Pb+Pb collisions}
\begin{figure}[!ht]
    \centering
    \includegraphics[width=.45\textwidth]{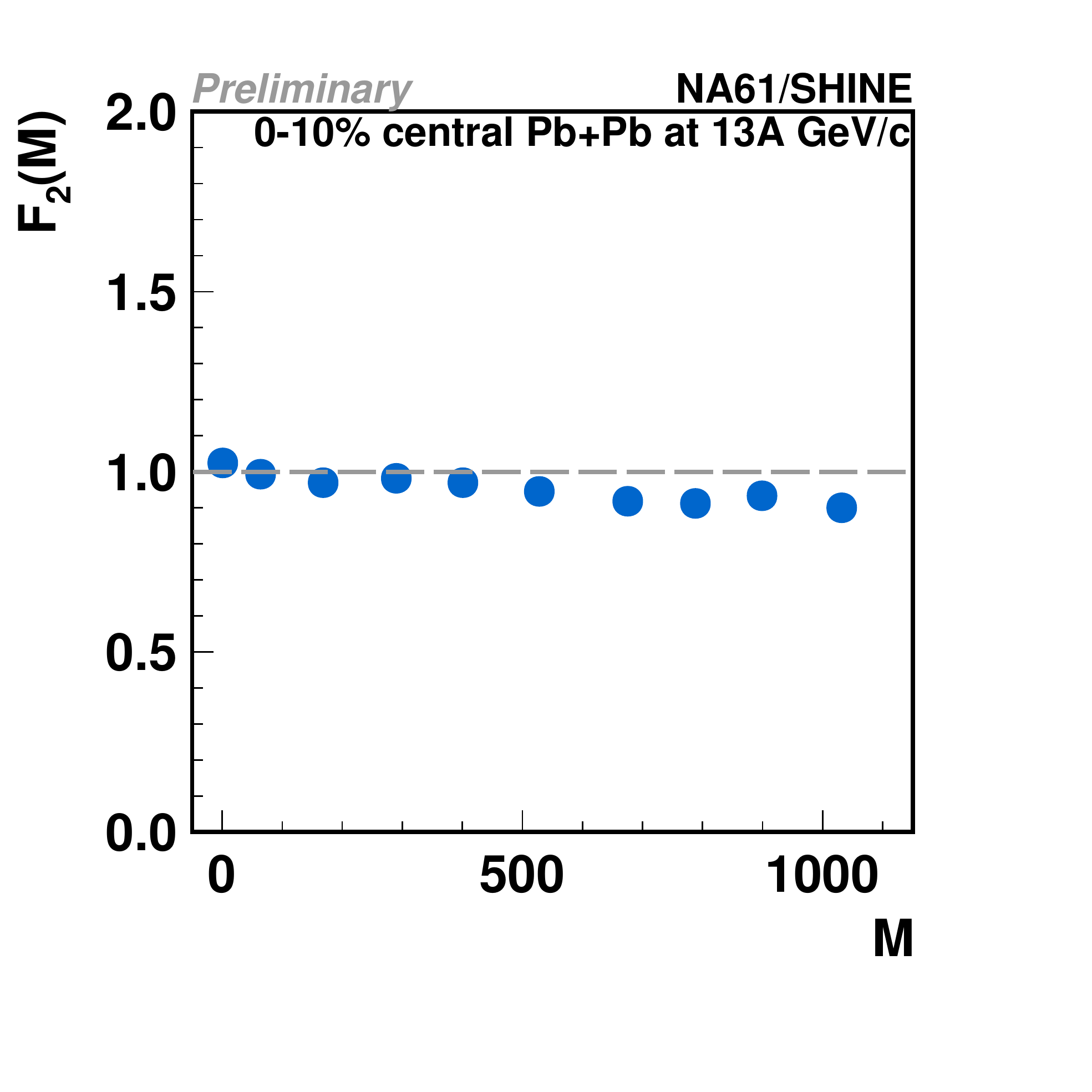}\quad
    \includegraphics[width=.45\textwidth]{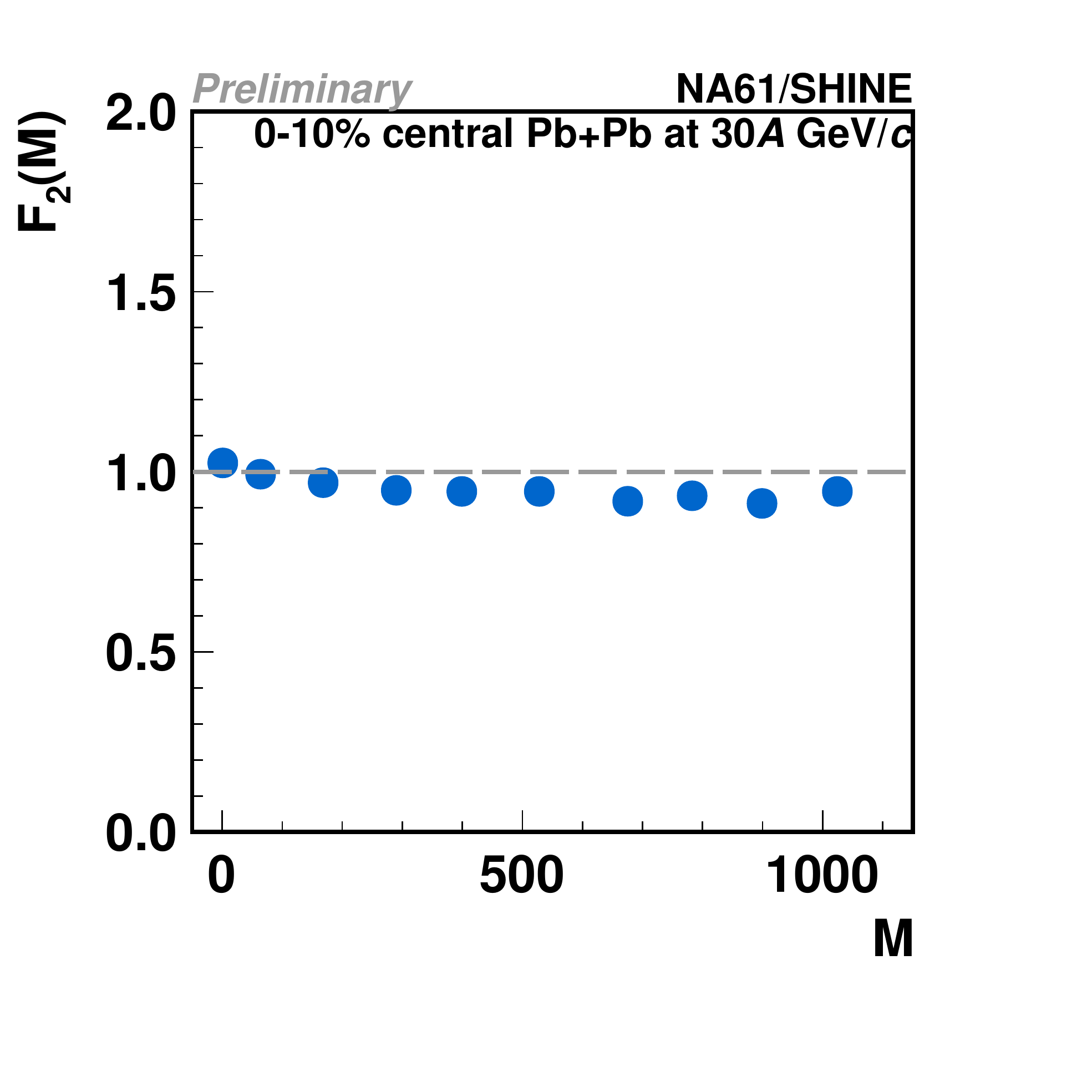}

    \caption{
        The dependence of the second scaled factorial moment of proton multiplicity distribution on the
        number of subdivisions in cumulative transverse-momentum space $M$ for $1 \leq M \leq 32^2$~\cite{Samanta:2021dxq} for central (0-10\%) \PbPb at 13\AGeVc (\emph{left}), central (0-10\%) \PbPb at 30\AGeVc collisions (\emph{right}).
        Only statistical uncertainties are indicated.
    }
    \label{fig:resultspbpb}
\end{figure}

Figures~\ref{fig:resultspbpb} present the dependence of the second-scaled factorial
moment for central (0-10\%) \PbPb at 13\AGeVc and central (0-10\%) \PbPb at 30\AGeVc collisions. 
The experimental results do not show any significant dependence on $M$.
There is no indication of the critical fluctuations for selected protons.

\subsection{Results for protons in central Ar+Sc collisions}
\begin{figure}[!ht]
    \centering
    \includegraphics[width=.33\textwidth]{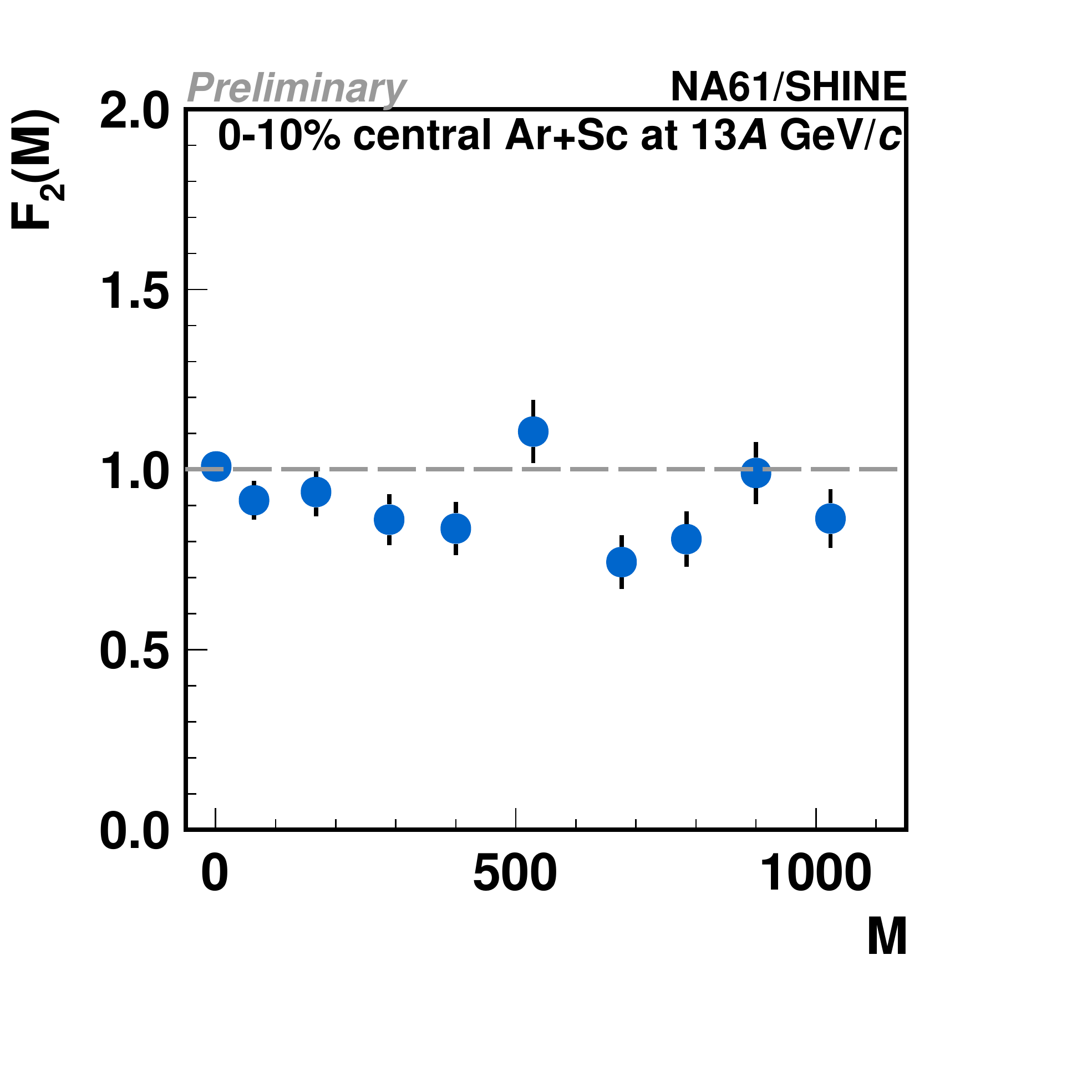}\hfill
    \includegraphics[width=.33\textwidth]{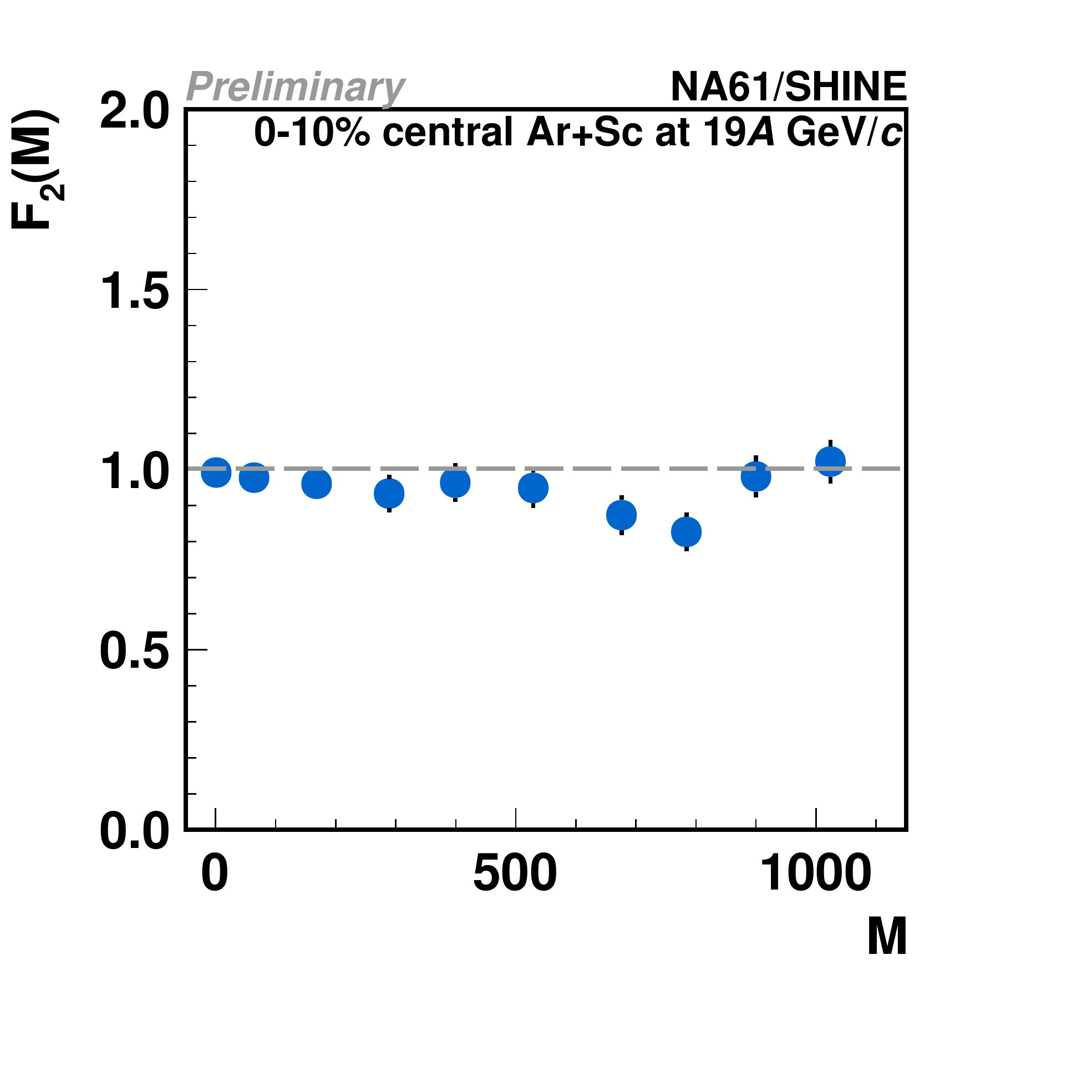}\hfill
    \includegraphics[width=.33\textwidth]{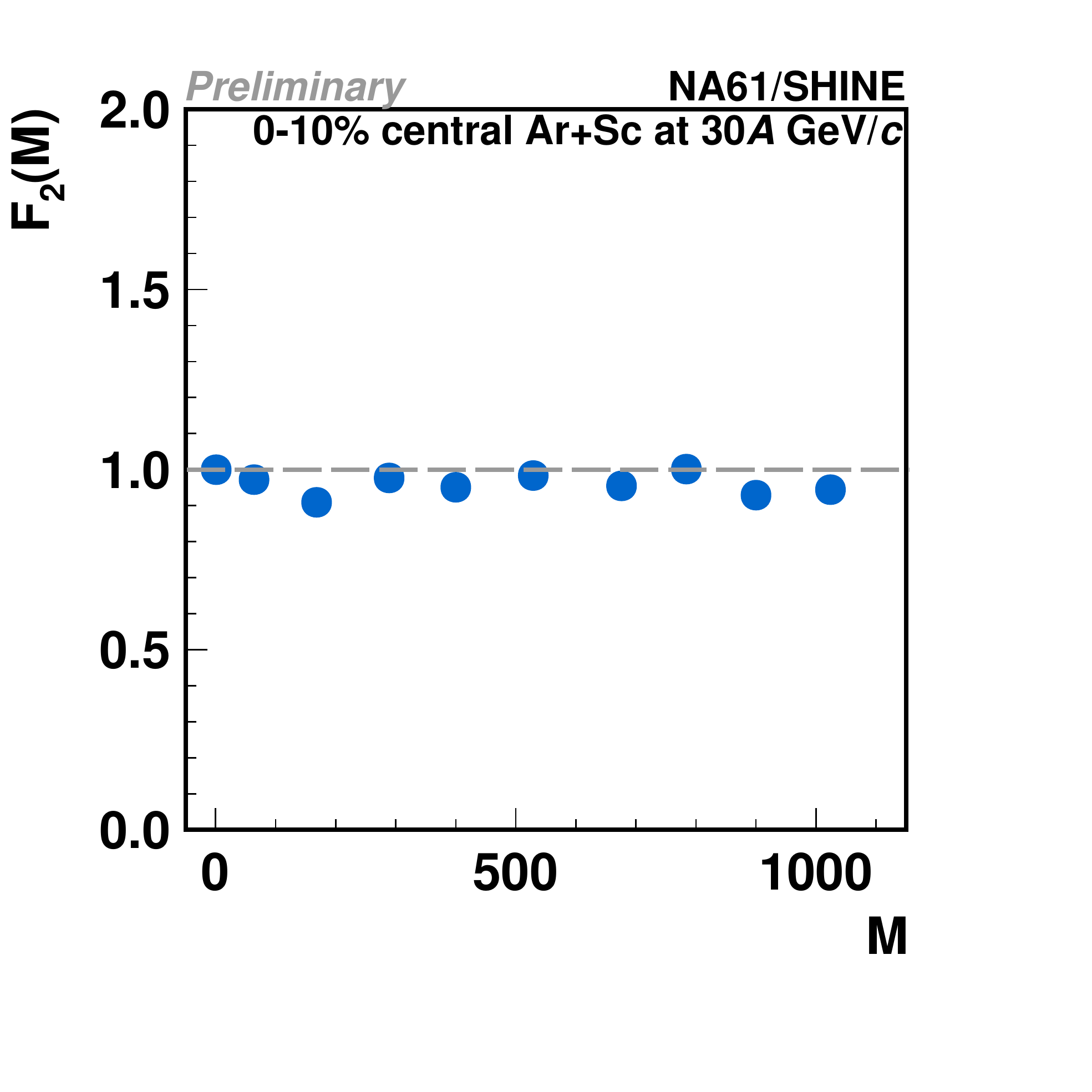}\\
    \includegraphics[width=.33\textwidth]{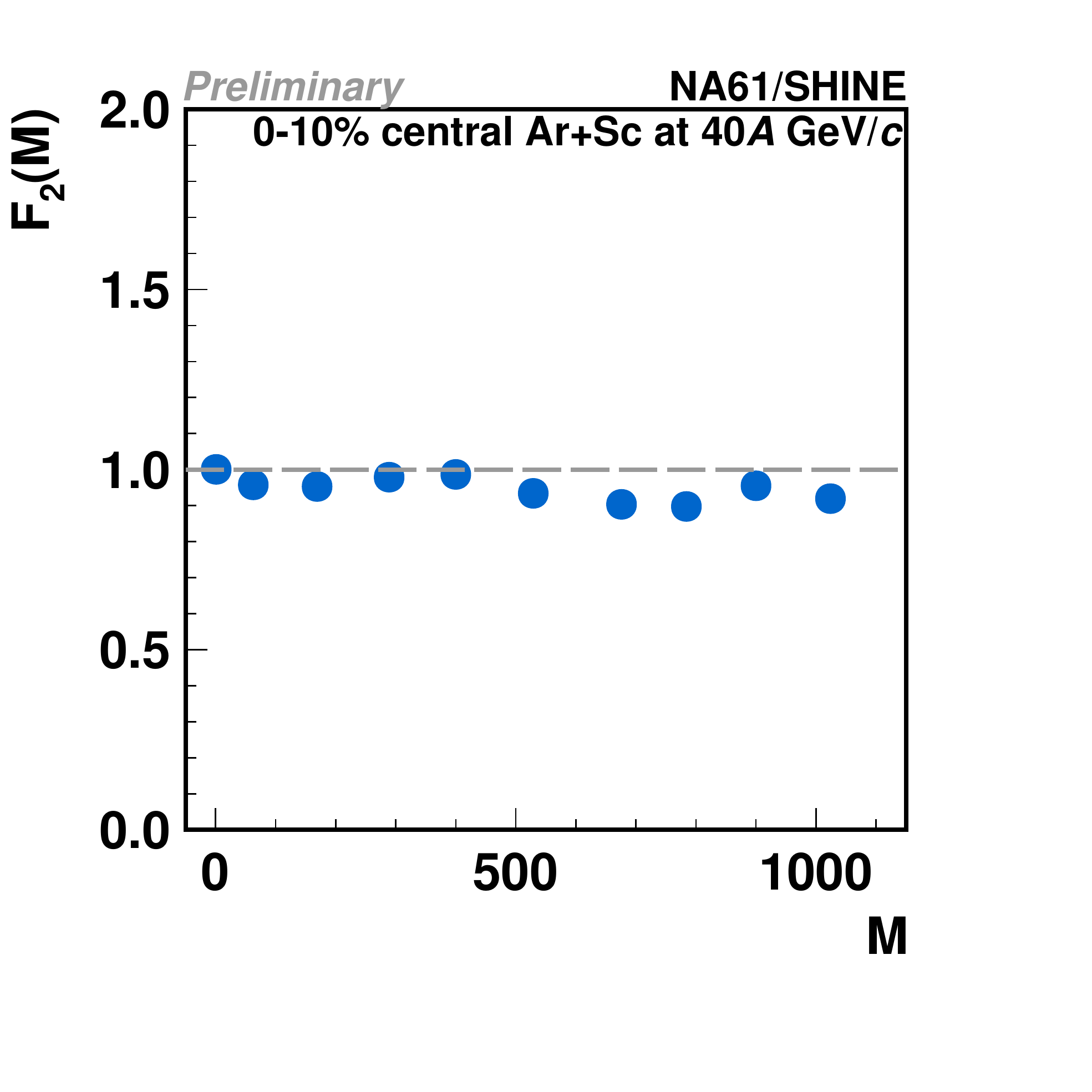}\hfill
    \includegraphics[width=.33\textwidth]{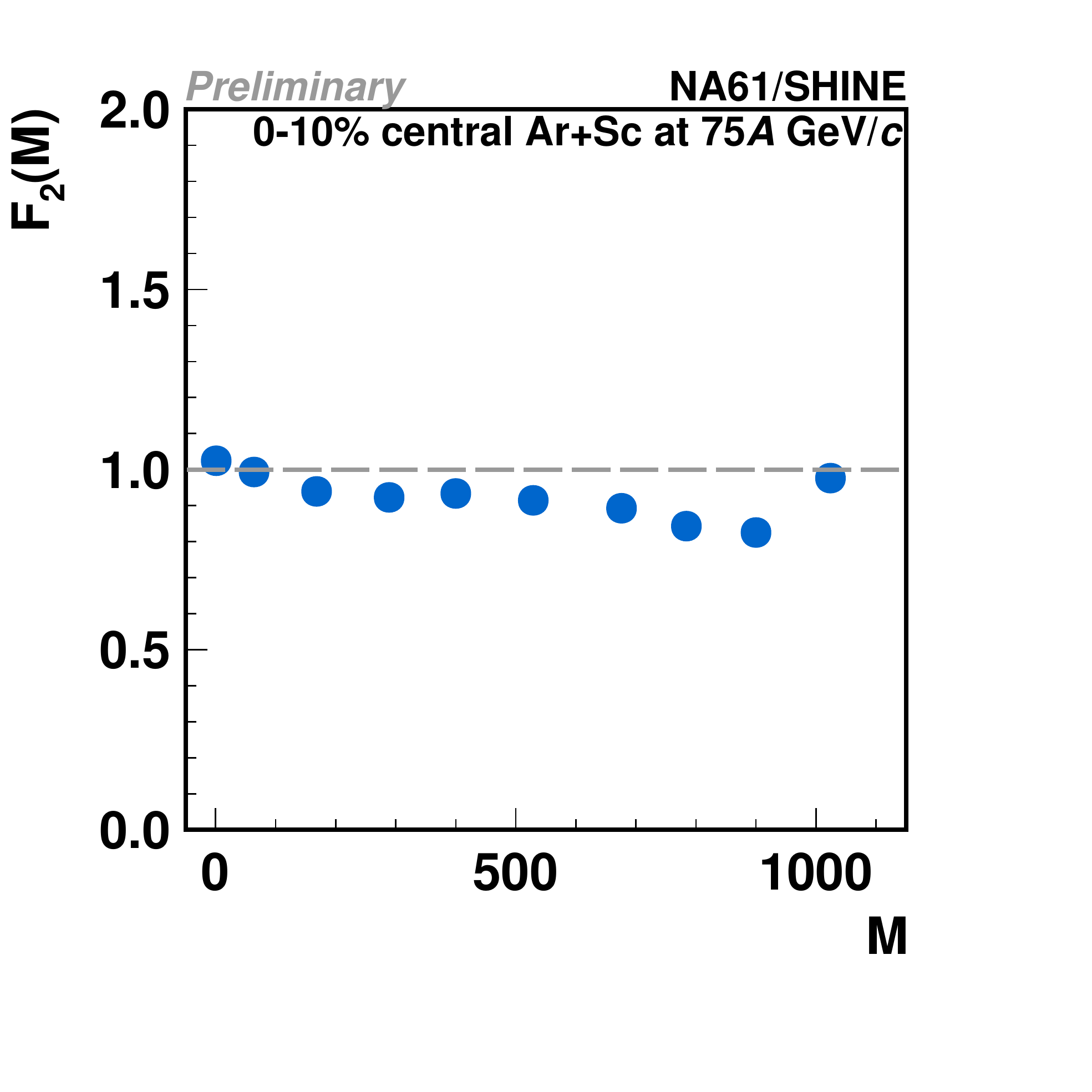}\hfill
    \includegraphics[width=.33\textwidth]{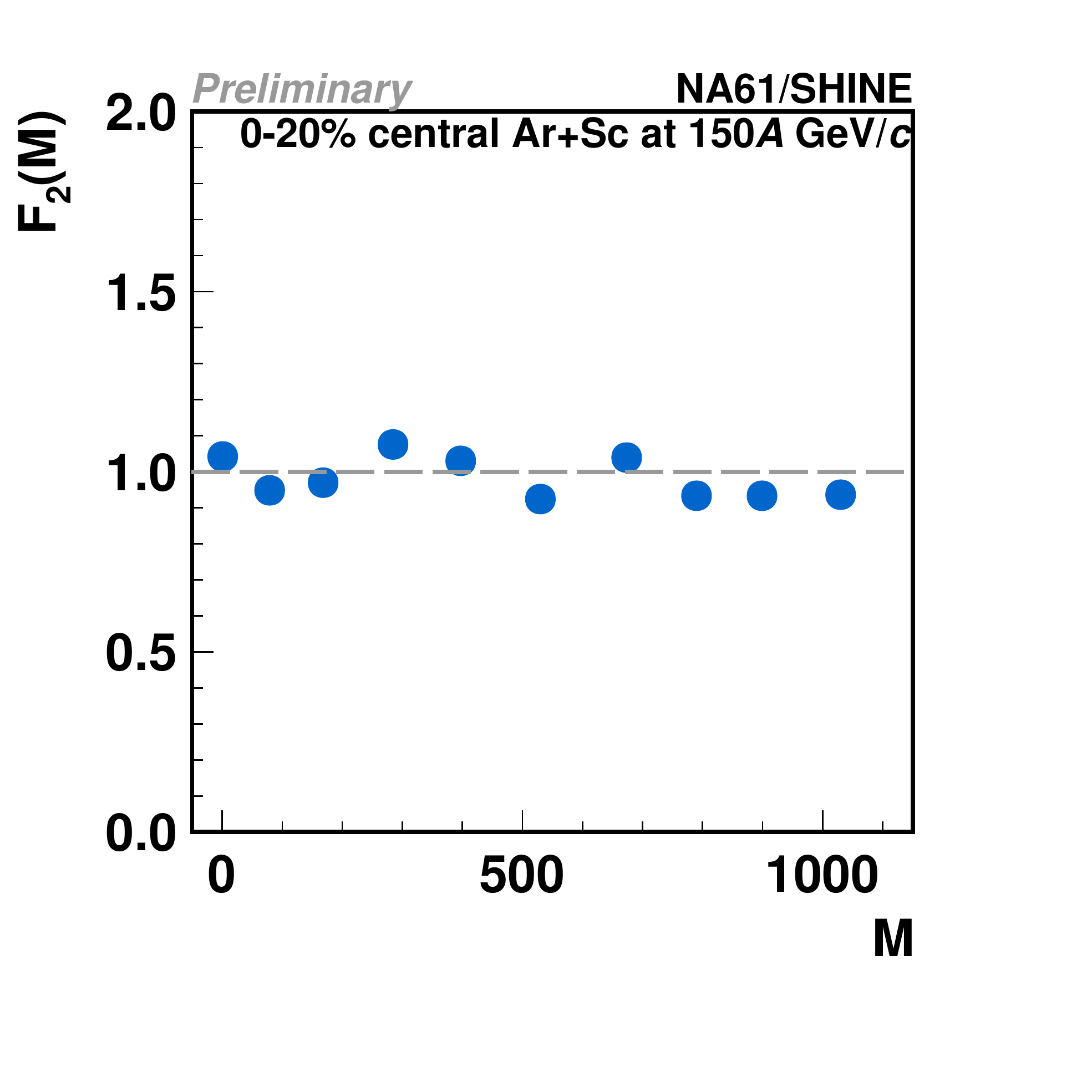}

    \caption{
        The dependence of the second scaled factorial moment of proton multiplicity distribution on the
        number of subdivisions in cumulative transverse-momentum space $M$ for $1 \leq M \leq 32^2$ for central (0-10\%) \ArSc at 13\A-75\AGeVc and central (0-20\%) \ArSc at 150\AGeVc collisions (from \emph{left} \emph{top} to \emph{bottom} \emph{right}).
        Only statistical uncertainties are indicated.
    }
    \label{fig:resultsarsc}
\end{figure}

Figures~\ref{fig:resultsarsc} present the dependence of the second-scaled factorial
moment for central (0-10\%) \ArSc at 13\A-75\AGeVc and central (0-20\%) \ArSc at 150\AGeVc collisions, 
The experimental results do not show any significant dependence on $M$.
There is no indication of the critical fluctuations for selected protons.

%% file: sections/models.tex

\subsection{Exclusion plots}
\label{sec:models}

\begin{figure}
    \centering
    \includegraphics[width=.33\textwidth]{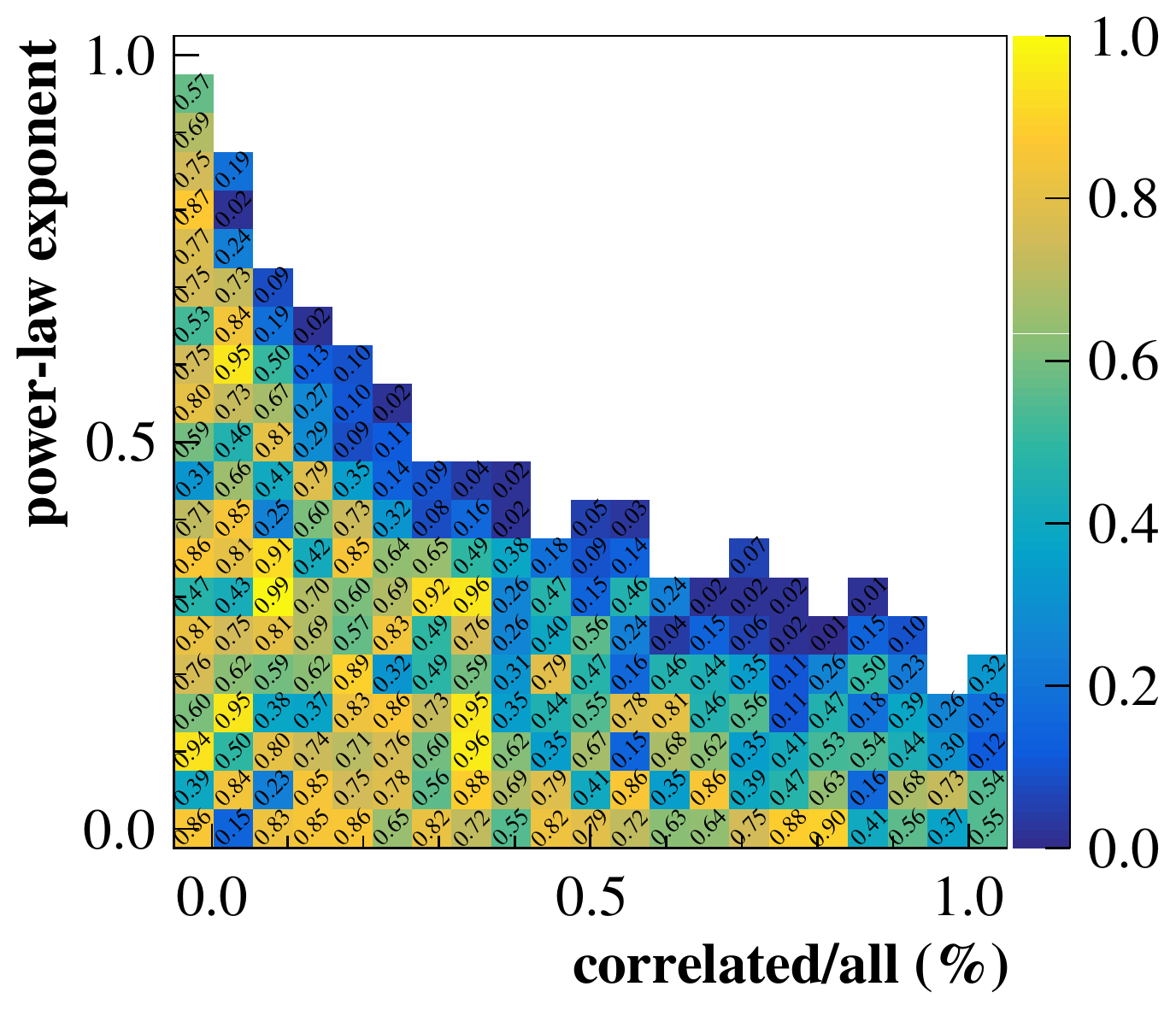}\hfill
        \includegraphics[width=.33\textwidth]{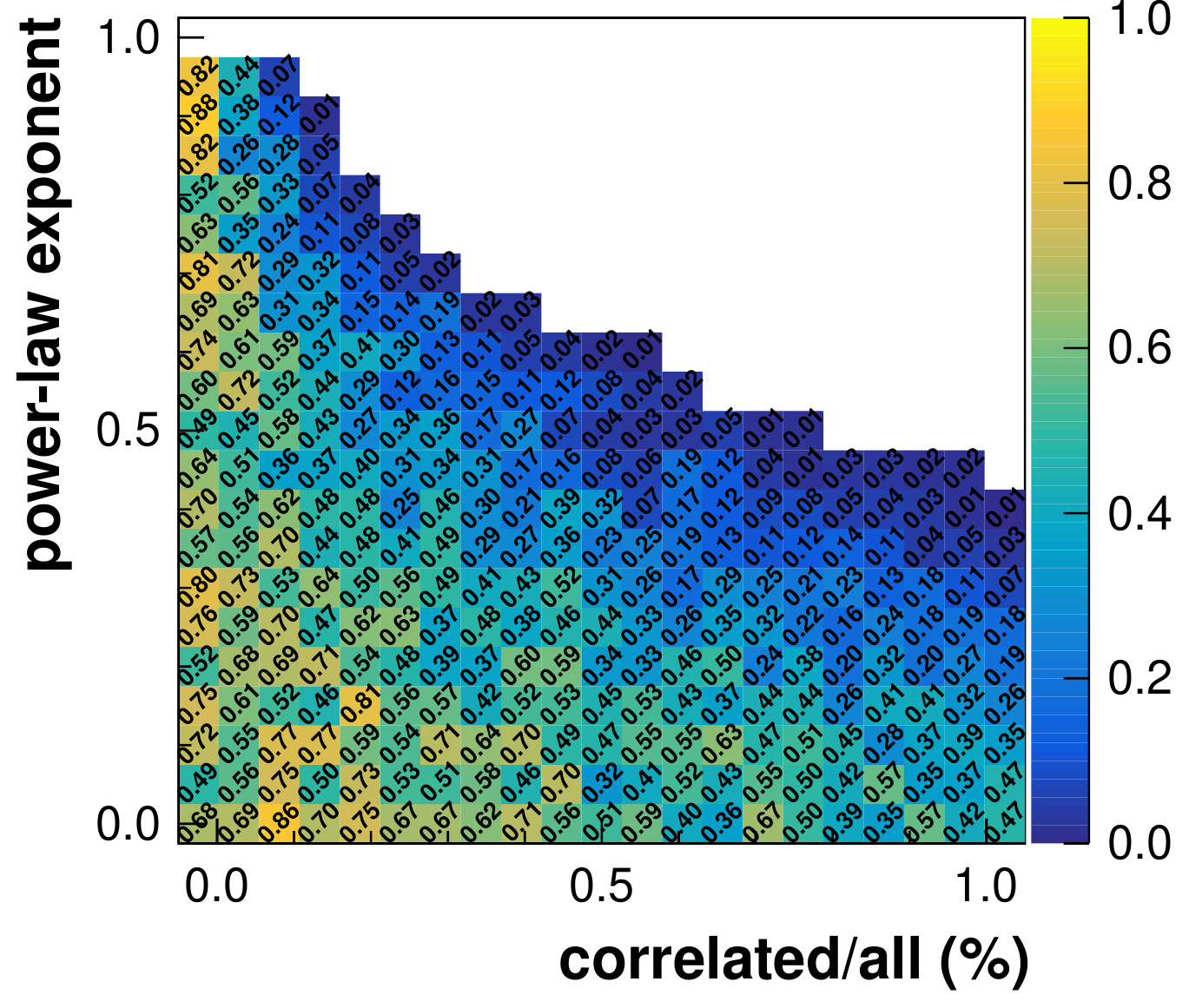}\hfill
    \includegraphics[width=.33\textwidth]{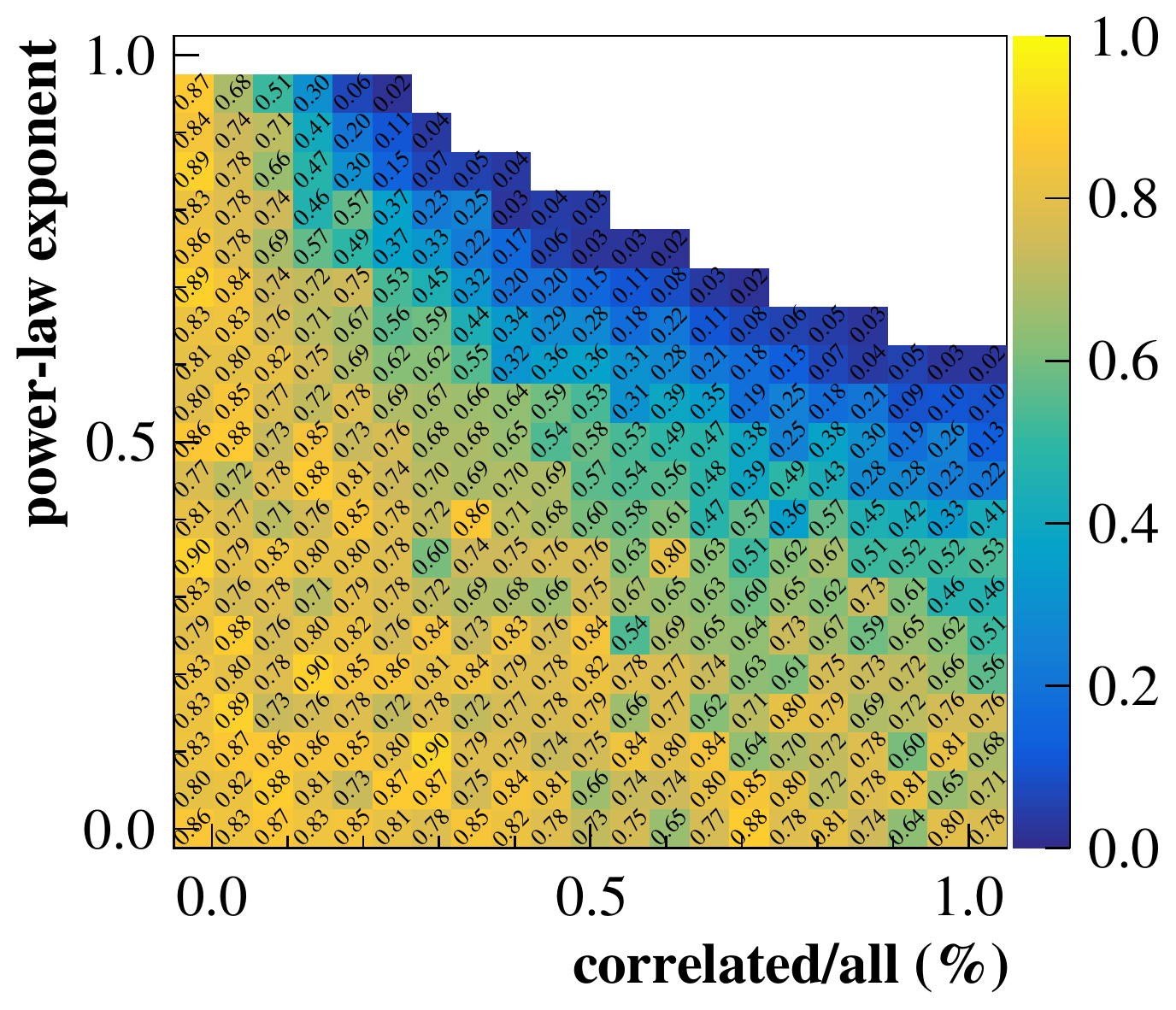}
    \caption{
        Exclusion plot, the p-values, for the Power-law Model parameters -- the fraction of correlated
        protons and the power-law exponent.
        The white areas correspond to  p-values less than 1\%.
        The exclusion plot for \PbPb at 13\AGeVc, \PbPb at 30\AGeVc and \ArSc at 150\AGeVc are shown from \emph{left} to \emph{right}. Exclusion plots for other data sets are expected soon.
    }
    \label{fig:exclusion-plot}
\end{figure}

This section presents exclusion plots for the parameters of the Power-law Model~\cite{Tobiasz:2022cpod} for a comparison of the experimental proton intermittency analysis results of central (0-10\%) \PbPb at 13\A, central (0-10\%) \PbPb at 30\AGeVc and central (0-20\%) \ArSc at 150\AGeVc collisions.\\
The model assumes that uncorrelated and correlated protons are produced with an equal to the measured single particle transverse momentum and multiplicity distribution. The model has two controllable
parameters:
\begin{enumerate}[(i)]
    \item fraction of correlated particles
    \item power-law exponent of the two-particle correlation function:
\end{enumerate}

$$
    \rho(|\Delta\overrightarrow{p_{T}}|) \sim |\Delta\overrightarrow{p_{T}}|^{-\phi_{2}}
$$
Many high statistics data sets are produced using the model.
Each data set has a different fraction of correlated particles (varying from 0 to 4\%) and a
different power-law exponent (varying from 0.00 to 0.95).\\
Next, all generated data sets have been analyzed the same
way as the experimental data. Obtained $F_{2}(M)$ results have been compared
with the corresponding experimental results and
$\chi^{2}$ and a p-value were calculated. 
Figure~\ref{fig:exclusion-plot} shows obtained p-values as a function of
the fraction of correlated protons and the power-law exponent.
White areas correspond to a p-value of less than 1\% and may be considered excluded.

%% file: sections/pion-intermittency.tex
\subsection{Results for negatively charged hadrons in central Pb+Pb collisions}
\begin{figure}[!ht]
    \centering
    \includegraphics[width=.33\textwidth]{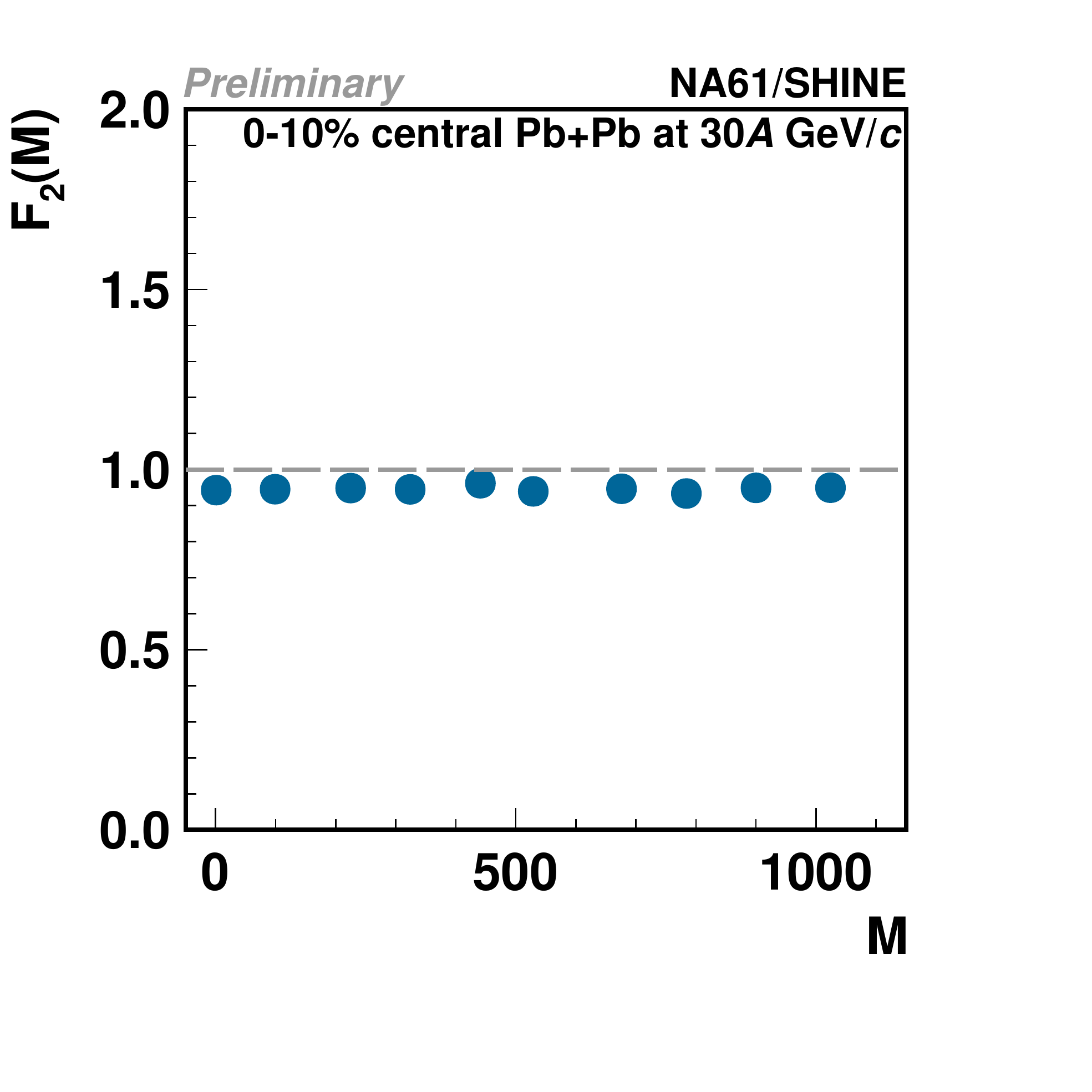}\hfill
    \includegraphics[width=.33\textwidth]{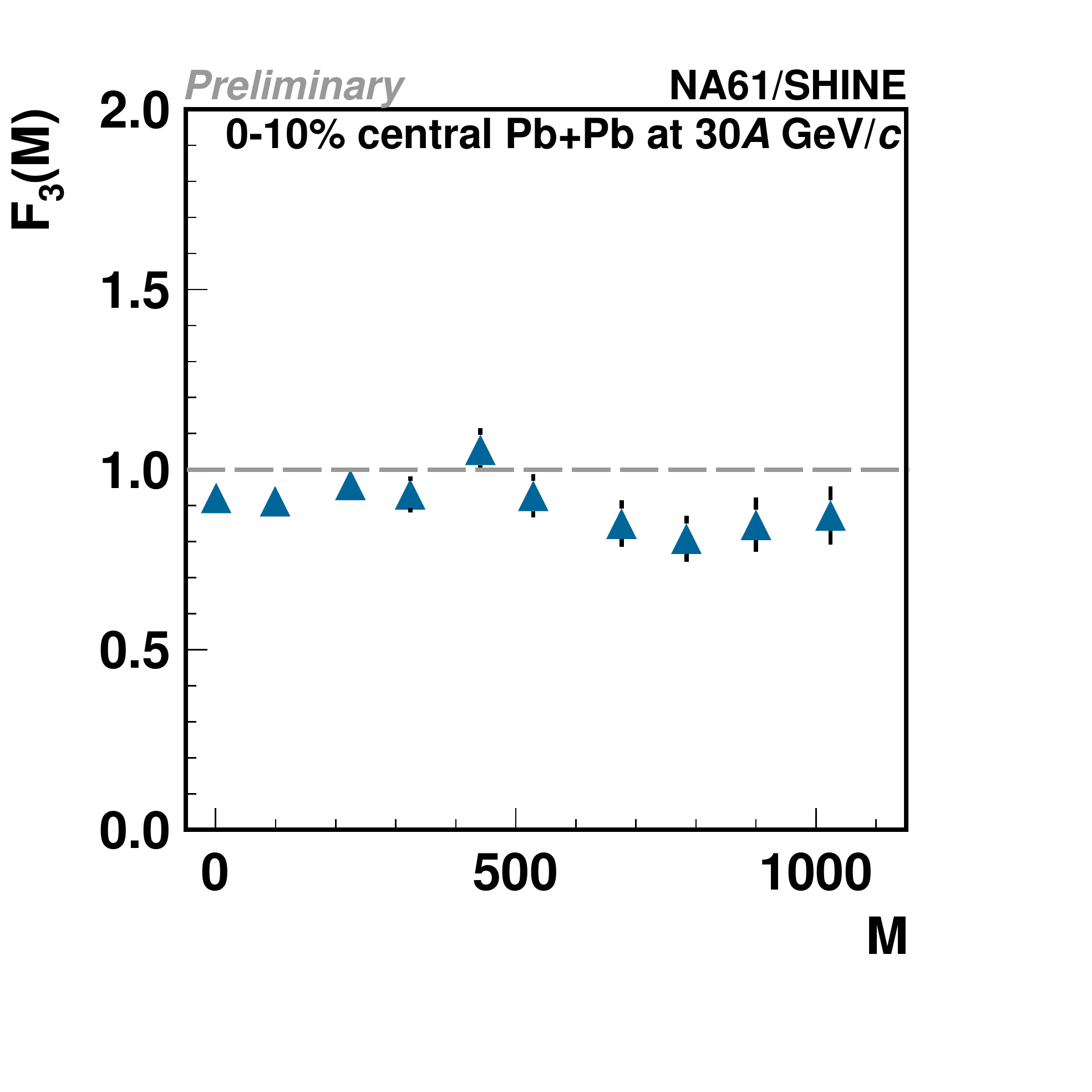}\hfill
    \includegraphics[width=.33\textwidth]{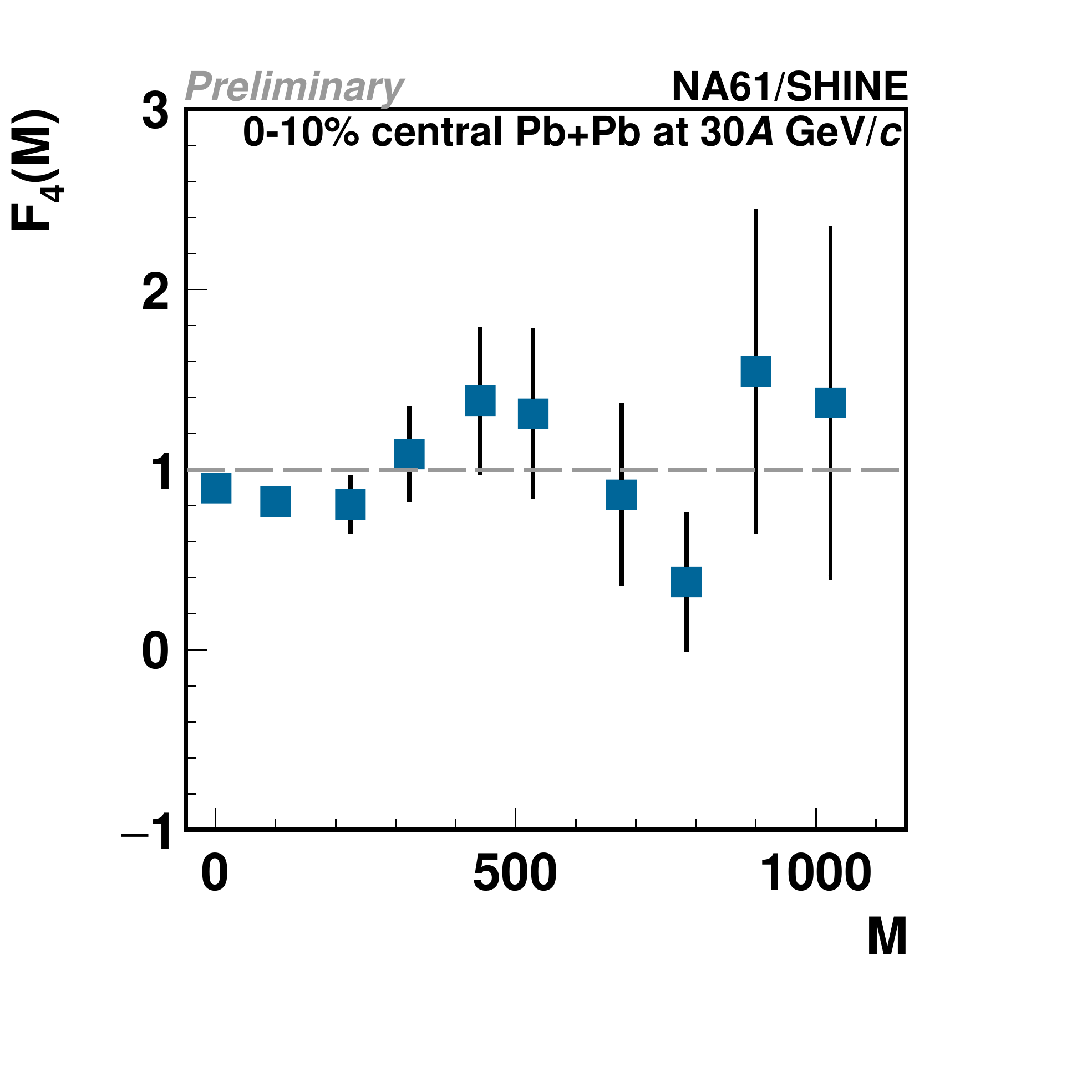}

    \caption{
        The dependence of the scaled factorial moments (F$_{2}$, F$_{3}$ and F$_{4}$) of negatively charged particle multiplicity distribution on the
        number of subdivisions in cumulative transverse-momentum space $M$ for $1 \leq M \leq 32^2$ for central (0-10\%) \PbPb at 30\AGeVc collisions. The results for F$_{2}$, F$_{3}$ and F$_{4}$ are plotted from \emph{left} to \emph{right}.
        Only statistical uncertainties are indicated.
    }
    \label{fig:results-coarse}
\end{figure}

Figures~\ref{fig:results-coarse} present the dependence of the scaled factorial
moments of multiplicity distribution of negatively charged hadrons on the number of subdivisions in cumulative transverse momentum space for central (0-10\%) \PbPb at 30\AGeVc collisions. For high event multiplicity of negatively charged hadrons, it is possible to calculate scaled factorial moments up to fourth-order moment (Eq.~\ref{eq:cp_1}). The experimental results do not show any significant dependence on $M$.
There is no indication of the critical fluctuations for selected negatively charged hadrons.

%% file: sections/summary.tex
\section{Summary}
\label{sec:summary}

This proceeding reports on the ongoing NA61/SHINE
search for the critical point of the strongly interacting matter. The results for central \PbPb collision at 13\AGeVc, 30\AGeVc and  \ArSc collisions at 13\A-150\AGeVc are presented.
The scaled factorial moments
of proton and negatively charged hadron multiplicity distribution at mid-rapidity
are shown as a function of  the number of subdivisions in the cumulative transverse-momentum components.
Independent data sets were used to calculate results for each subdivision.
The results show no indication of the intermittency signal.
The intermittency analysis of other reactions recorded within the \NASixtyOne
program on strong interactions is well advanced, and new results should
be expected soon.
\begin{figure}[!ht]
    \centering
    \includegraphics[width=.525\textwidth]{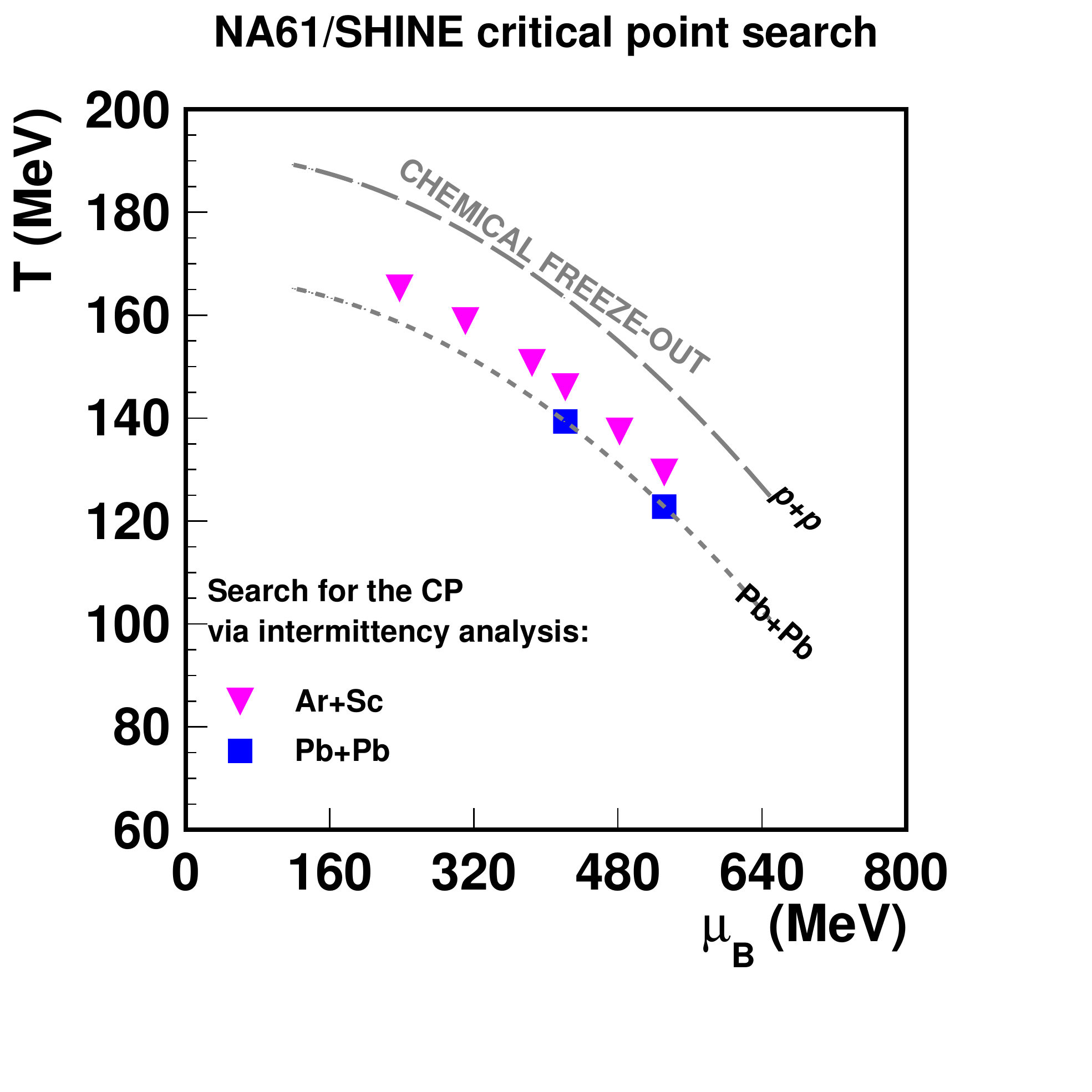}

    \caption{
    %
Diagram of chemical freeze-out temperature and chemical potential. The dashed line indicates parameters in \pp interactions and  the dotted line in central Pb+Pb collisions~\cite{Becattini:2005xt}. The colored points mark reactions in the $T-\mu_{B}$ phase diagram for which search for the critical point was conducted.
    }
    \label{fig:summary}
\end{figure}